\documentclass[lettersize, journal]{IEEEtran}
\usepackage{cite}
\usepackage{array}
\usepackage{amsmath,amssymb,amsfonts}
\usepackage{algorithmic}
\usepackage{graphicx}
\usepackage{gensymb}
\usepackage{textcomp}
\usepackage{float}
\usepackage{dblfloatfix}
\usepackage[caption=false,font=normalsize,labelfont=sf,textfont=sf]{subfig}
\usepackage{url}
\usepackage{verbatim}
\begin{document}
\title{Niobate-on-Niobate Resonators with Aluminum Electrodes}
\author{Yiyang Feng, \IEEEmembership{Student Member, IEEE}, Sen Dai,  \IEEEmembership{Member, IEEE}, Sunil Bhave, \IEEEmembership{Senior Member, IEEE}
\thanks{
\\
(\textit{Corresponding author: Sunil A. Bhave}) }
\thanks{Yiyang Feng is with Department of Physics and Astronomy, Purdue University, West Lafayette, IN 47907, USA.}
\thanks{Sen Dai is with Qorvo, Inc., Apopka, FL 32703, USA. }.
\thanks{Sunil A. Bhave is with the School of Electrical and Computer Engineering, Purdue University, West Lafayette, IN 47907, USA.}
}

\maketitle

\begin{abstract}
In this work, we have successfully engineered and examined suspended laterally vibrating resonators (LVRs) on a lithium niobate thin film on lithium niobate carrier wafer (LN-on-LN) platform, powered by aluminum interdigital transducers (IDTs). Unlike the lithium niobate-on-silicon system, the LN-on-LN platform delivers a stress-neutral lithium niobate thin film exhibiting the quality of bulk single crystal. The creation of these aluminum-IDTs-driven LN-on-LN resonators was achieved utilizing cutting-edge vapor-HF release techniques. Our testing revealed both symmetric (S0) and sheer horizontal (SH0) lateral vibrations in the LVR resonators. The resonators displayed a quality factor (Q) ranging between 500 and 2600, and coupling coefficient (\( \mathbf{k_{eff}^2} \)) up to 13.9\%. The figure of merit (FOM) \(\mathbf{k_{eff}^2 \times Q}\) can reach as high as 294. The yield of these devices proved to be impressively reliable. Remarkably, our LN-on-LN devices demonstrated a consistently stable temperature coefficient of frequency (TCF) and good power handling. Given the low thermal conductivity of lithium niobate, our LN-on-LN technology presents promising potential for future applications such as highly sensitive uncooled sensors using monolithic chip integrated resonator arrays.
\end{abstract}


\section{Introduction}
\label{sec:introduction}

Emerging 5G technology has created enormous opportunities and challenges in telecommunication industry. In order to cater to burgeoning demands for ultra-fast data transfer rate and a vast capacity of large-scale machine communications in the 5G wireless sector, there is a pressing need for high-performance, multi-frequency duplexer and multiplexer devices with an elevated coupling coefficient \(\mathrm{k_{eff}^2}\) and quality factor Q. Aluminum Nitride (AlN) Film Bulk Acoustic Resonators (FBAR) and Contour-Mode Resonators (CMR) have been proposed to meet the criteria from RF industry. AlN FBAR technology has shown notable promise with high quality factors (Q) and a considerable fractional bandwidth of up to 7\%\cite{6174178}. Nevertheless, its capacity for incorporating multiple frequencies on a single chip remains constrained, primarily because the resonant frequency of the FBAR is dictated by the thickness of the thin film. Conversely, AlN CMRs present a different challenge, as their coupling coefficient (\(\mathrm{k_{eff}^2}\)) falls short, registering below 2\%\cite{5361520}. 

\begin{figure}[t]
\centerline{\includegraphics[width=\columnwidth]{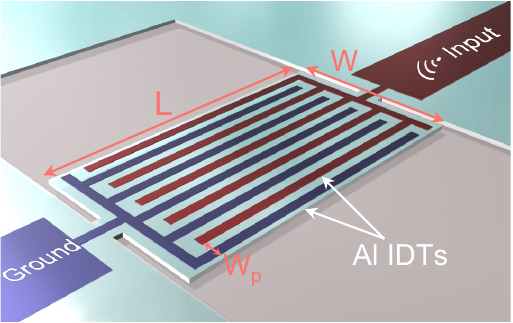}}
\caption{3D model of LN-on-LN Laterally vibrating resonators. Red and blue indicate the IDT fingers connected to the source and ground, respectively. }
\label{fig:1}
\end{figure}

\begin{figure}[h]
\centerline{\includegraphics[width=\columnwidth]{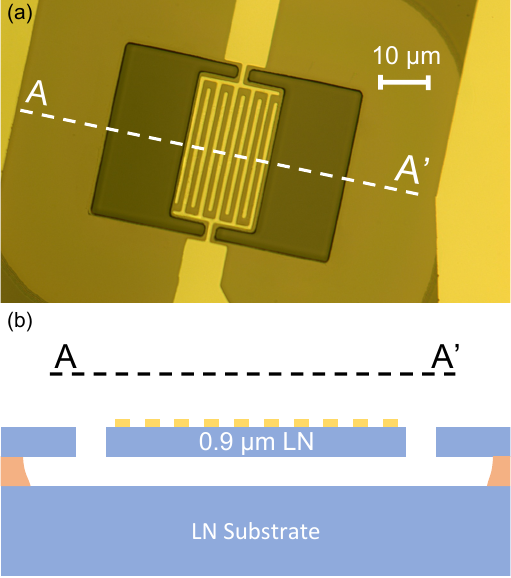}}
\caption{
    \textbf{(a)} Optical image of LN-on-LN LVR devices. 
    \textbf{(b)} Cross sectional view of released devices along A-A' dashed line in subfigure (a).
}
\label{fig:2}
\end{figure}

Recent efforts have been dedicated to the lithium niobate (LN) platform. In contrast to AlN devices, the large \(\mathrm{d_{31}}\), \(\mathrm{d_{15}}\) and \(\mathrm{d_{26}}\) piezoelectric coefficients enable very high coupling coefficient and figure of merit (FOM) \(\mathrm{k_{eff}^2 \times Q}\) within the lithium niobate devices\cite{Xu:2020aa, 6619408, 7265019, TAKEUCHI2002463, olsson2014lamb, 6619408, 7863565, OLSSON2014183, Gruenke:2023pas}. Despite its strong potential, future applications on lithium niobate technology face the following challenges. Although conventional lithium niobate SAW resonators are built on top of silicon carrier wafer, the large lattice constant and thermal expansion coefficient mismatch between LN and Si make the fabrication of free-standing BAW resonators with anchor suspension thermally unstable\cite{9385101, olsson2014lamb, 6381492}. Earlier studies have leveraged the LN-on-LN platform as a strategy to address these limitations\cite{6619408, 7265019, TAKEUCHI2002463, 7005450, riedel2023efficient}. Nevertheless, these design approaches require the use of high-damping metal electrodes due to compatibility concerns during the fabrication process. This requirement, in turn, compromises the quality factor (Q).

\begin{figure}[th]
\centerline{\includegraphics[width=\columnwidth]{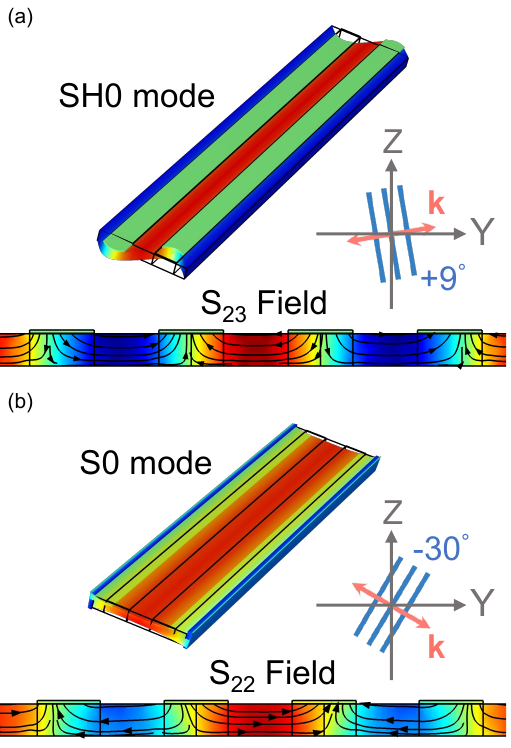}}
\caption{
    \textbf{(a)} Mode shape of SH0 mode. 
    \textbf{(b)} Mode shape of S0 mode.
}
\label{fig:3}
\end{figure}

In this work, we present our latest progress in the LN-on-LN bulk acoustic wave (BAW) technology with aluminum electrodes. The bulk-quality single crystal LN thin film is integrated through thermal matched bonding and polishing to create stress-free bonding. With a newly invented fabrication process, we successfully create a LN LVR device with low damping aluminum electrodes. By varying the orientation of the resonator, SH0 modes and S0 modes can be selectively excited on the same chip with Q ranging from 800 to 2500 and \(\mathrm{k_{eff}^2}\) up to 13.9\%. The highest figure of merit (FOM) \(\mathrm{k_{eff}^2 \times Q}\) of our device can reach 294, which is one of the highest in the reports. Moreover, our devices also demonstrate a stable temperature coefficient (TCF) as high as -100.1 ppm/K for SH0 modes and -65.3 ppm/K for S0 modes at various power levels within the resonator’s linear regime. The capability to operate at different temperature and power level reveals its potential for application such as a highly sensitive uncooled bolometer.

\section{Design and Modeling}
\label{sec:model}

\subsection{Design properties}
The LN-on-LN LVR devices that we have developed feature an anchor-supported suspended piezoelectric thin film. A 0.9 \(\mathrm{\mu m}\) thick stress free X-cut lithium niobate thin film is created on top of 1 \(\mathrm{\mu m}\) oxide buffer layer. Oxide layer also serves as sacrificial layer to release the piezoelectric film. The suspended structure is designed with a length of L = 70 \(\mathrm{\mu m}\) and a width of W = 44 \(\mathrm{\mu m}\). This is integrated with meticulously positioned aluminum interdigital transducers (IDTs), aligned in a set of specific in-plane orientations to stimulate various acoustic modes. For symmetric configuration, IDTs consist of M = 3, 5, 7, 9, 11 electrodes, while for anti-symmetric configuration, IDTs contain M = 10 electrodes. The symmetric and anti-symmetic configuration will affect the coupling of different mode families respectively. Pitch \(\mathrm{W_p}\) is chosen to be 4 \(\mathrm{\mu m}\) and metallization ratio is set to be 50\%. Compared with gold, which is notorious for its high internal friction loss, aluminum offers a lower intrinsic loss and a superior acoustic impedance match to lithium niobate. As a result, it has seen extensive use in LN-on-Si devices\cite{5361520, 8439516}. With novel fabrication technology, aluminum IDTs are introduced to our LN-on-LN platform. 

\subsection{Finite Element Modeling}
We carried out 3D finite element analysis using COMSOL on lithium niobate resonators with varying orientations. For SAW wave devices, the frequency response has a sinc dependce on IDT pitch\cite{7005450}. In our case, where lamb waves are launched within a laterally vibrating resonator, the frequency response is complicated by the Fabry-Perot (FP) cavity formed by the vertical sidewall. The mechanical boundary conditions of FP cavity enforces more constraints, where the \(\mathrm{N^{th}}\) order lamb wave with a wavelength \(\mathrm{\lambda}\) satisfying \(\mathrm{N \times \lambda/2 = W}\) will emerge. However, only the waves with best match to IDT configuration have good coupling and high Q. Through simulations, it has been deduced that escalating the number of IDT fingers (denoted as M) facilitates enhanced coupling of the resonator. Upon reaching a state where the product of M and the IDT pitch (\(\mathrm{W_p}\)) is equal to the overall device width (W), an optimal coupling scenario is realized. Consequently, the wave with an order defined by (N = M - 1) exhibits the highest degree of coupling efficiency in the system. Notably, with varied orientation of IDTs, two types of lamb wave modes, SH0 and S0 , are detected in the simulation. When IDTs are aligned +9 degree relative to the +z axis of X-cut LN thin film, where ``+" sign denotes counterclockwise rotation, the coupling of SH0 modes is optimized. The resultant acoustic waves travel at a +9 degree angle to the +y axis with SH0 modes of different orders spanning 300 MHz to 500 MHz. As illustrated in Fig. \ref{fig:3}, SH0 modes predominantly feature \(\mathrm{S_{23}}\) stress field component, signifying that \(\mathrm{d_{15}}\) and \(\mathrm{d_{26}}\) piezo coefficients are playing substantial roles in the excitation of SH0 modes. In comparison, when IDTs are aligned at a -30 degree to the +z axis (a ``-" sign signifying clockwise rotation), different orders of S0 modes materialize between 700 MHz and 800 MHz with maximized coupling. In the case of S0 mode, the \(\mathrm{S_{22}}\) component dominates stress field, attributed to \(\mathrm{d_{22}}\) piezo coefficient.

\section{Fabrication}
\label{sec:fabrication}

\subsection{Process Flow}

\begin{figure}[t]
\centerline{\includegraphics[width=\columnwidth]{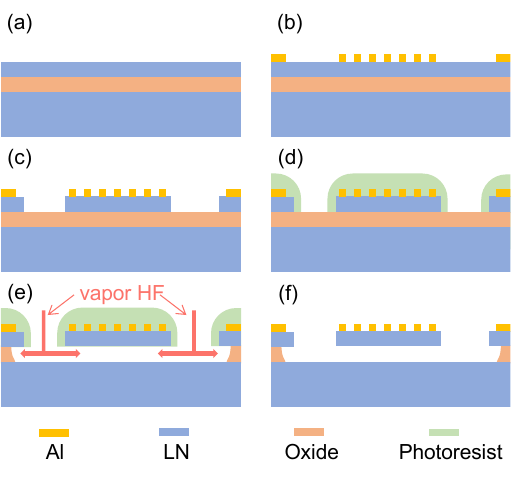}}
\caption{
    \textbf{(a)} Sample is prepared via bonding 0.9 \(\mathrm{\mu m}\) stress free X-cut LN niobate onto 1 \(\mathrm{\mu m}\) oxide buffer layer on X-cut LN carrier wafer. 
    \textbf{(b)} Aluminum electrodes is patterned through lift-off process.
    \textbf{(c)} Ion mill etching on LN thin film is performed to define the device geometry.
    \textbf{(d)} Photoresist layer is patterned and reflowed with promoted adhesion.
    \textbf{(e)} Vapor HF flow applied onto the sample at elevated temperature to remove oxide layer without damaging Al electrodes.
    \textbf{(f)} The sample is transferred to acetone to remove photoresist. Another drying at critical point of \(\mathrm{LCO_2}\) is conducted to fully release the structure.
}
\label{fig:4}
\end{figure}

\begin{figure}[t]
\centerline{\includegraphics[width=\columnwidth]{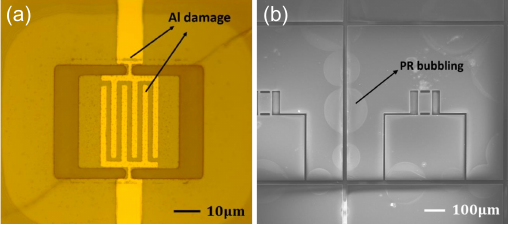}}
\caption{ 
\textbf{(a)} Image indicating undesirable undercut and damage in Al electrodes in the absence of adequate masking.
\textbf{(b)} SEMs image showing bubbling and peeling of photoresist without adhesion promotion
}
\label{fig:add_1}
\end{figure}

The preparation of LN-on-LN sample entails precise alignment and bonding of lithium niobate samples with 1 \(\mathrm{\mu m}\) oxide buffer layer to alleviate stress. The upper niobate layer is subsequently thinned to 0.9 \(\mathrm{\mu m}\) using mechanical polishing. Following the sample preparation, we deposit 100 nm thick aluminum electrodes through evaporation and lift-off procedure. The ensuing step involves patterning the LN structure. Conventionally, lithium niobate is defined via fluorine-based reactive ion etching (RIE). This tends to result in sloppy sidewall, rough etching surface and significant redeposition of \(\mathrm{LiF_2}\)\cite{6381492, benchabane2009highly}. To circumvent these issues, we utilize an ion mill procedure\cite{7005450, wang2016multi} under argon plasma. The argon ions are directed onto our sample at a steep angle at 14\degree{} to etch through the niobate layer at a rate of 56 nm/min. A subsequent 70\degree{} shallow angle ion mill is performed to clean up the sample and remove any redepositted materials left by ion mill process. After completing the ion milling process, the sample is immersed in acetone, and a 5-minute sonication is carried out to remove the photoresist. 

\begin{figure}[t]
\centerline{\includegraphics[width=\columnwidth]{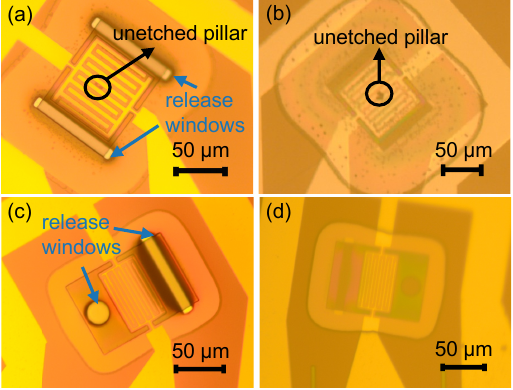}}
\caption{ 
\textbf{(a)} Front side image showing the release structure with symmetric release windows. Unetched pillar can be found at the center behind IDTs. 
\textbf{(b)} Back side image showing the release structure with symmetric release windows. Unetched pillar can be found at the center. 
\textbf{(c)} Front side image showing the release structure with asymmetric release windows. No Unetched pillar can be found at the center behind IDTs. 
\textbf{(d)} Back side image showing the release structure with asymmetric release windows. No Unetched pillar can be found at the center. 
}
\label{fig:add_1}
\end{figure}

\begin{figure}[t]
\centerline{\includegraphics[width=\columnwidth]{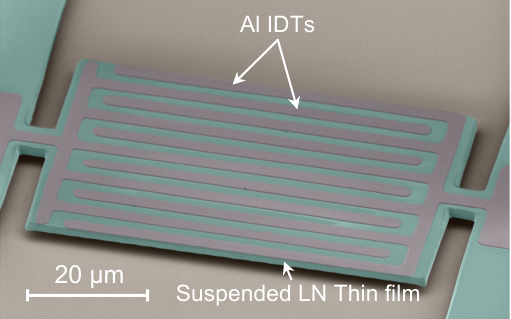}}
\caption{ False color SEM of the LN-on-LN laterally vibrating resonators.
}
\label{fig:5}
\end{figure}

\begin{table}[t]
\caption{Comparison of different method to release niobate structure}
\label{table}
\setlength{\tabcolsep}{3pt}
\begin{tabular}{>{\centering}p{84pt}>{\centering}p{90pt}>{\centering\arraybackslash}p{50pt}}
\hline\hline
Methods &  
Fabrication Results & 
RF Response \\
\hline
No adhesion promotion & 
Micro-damage on Al, & 
 \\
+ Symmetric windows & 
partially released structure & 
Broken \\
\hline
Surpass 4000 treatment & 
No damage on electrodes, & 
 \\
+ Symmetric windows & 
partially released structure & 
Poor \\
\hline
Surpass 4000 treatment & 
No damage on electrodes, & 
 \\
+ Asymmetric windows & 
fully released structure & 
Good \\
\hline\hline
\end{tabular}
\label{tab:1}
\end{table}

\subsection{Protection of Aluminum Electrodes}

The pivotal step in our process flow is the protection of aluminum IDTs while releasing the niobate structure. We opt for vapor HF technology over traditional buffered oxide etch (BOE) technology, largely due to the incompatibility between aluminum IDTs and BOE solution. The latter can easily penetrate the mask through the pinholes in the photoresist layers, leading to erosion of the aluminum IDTs. Conversely, vapor HF does not have the capacity to etch aluminum in the absence of water, making it a suitable candidate for processes involving aluminum IDTs. Nevertheless, during our experiments, we observed that moisture could accumulate on the sample surface, exacerbating the etching effect of vapor HF when aluminum IDTs were exposed. Therefore, it is imperative to employ effective masking strategies to mitigate moisture accumulation. Prior to spinning the photoresist mask layer, the surface is treated with Surpass 4000 resist, which activates the surface and enhances adhesion. The presence of Surpass 4000 adhesion is crucial. It maintains the integrity of photoresist layer under vapor HF ambience. Without adhesion promotion, vapor HF tends to yield highly undesirable bubbling or leakages inside the photoresist layer, where HF and moisture accumulate and cause damages to alumiunum IDTs. Surpass 4000 can also ensure better mask coverage on the sidewall. Once the surface is treated, we then pattern a 7 \(\mathrm{\mu m}\) photoresist layer on treated surface, followed by a hard bake at 110 \degree C. This is performed to reflow the photoresist, removing moisture and cure pinholes within the layer.   

\subsection{Release of Niobate Structure}

\begin{figure}[t]
\centerline{\includegraphics[width=\columnwidth]{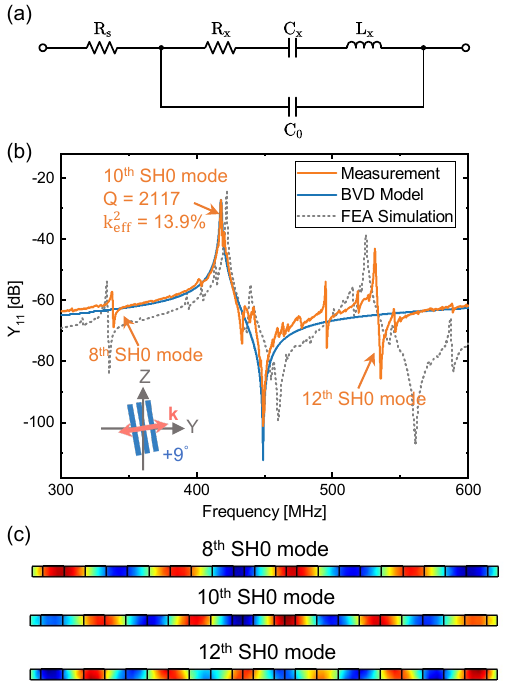}}
\caption{ 
\textbf{(a)} BVD circuit model applied to analyze the RF behaviors of our resonators.
\textbf{(b)} Measured results, BVD model analysis and finite  element analysis of SH0 modes with +9\degree{} IDT orientation. 11 IDT fingers are equipped. Different orders of SH0 modes are marked in the spectrum.
\textbf{(c)} Cross sectional view of stress field for \(\mathrm{8^{th}}\), \(\mathrm{10^{th}}\) and \(\mathrm{12^{th}}\) order SH0 modes. blue and red indicates \(\mathrm{S_{23}}\) field with different signs.
}
\label{fig:6}
\end{figure}

Once the patterning of photoresist is completed, the sample is transferred to an electrostatic holding chuck within a commercial HF-vapor etcher. In this study, the release windows are intentionally designed to have an asymmetric configuration. Previous attempts employing symmetric release windows resulted in partial release of the structures, leaving unetched pillars at the central region. This is attributed to the oxide from the peripheral areas consuming substantial amounts of HF, consequently impacting the dynamics of the HF flow. At the symmetric point, i.e., the center, the HF flow is diminished, leading to a reduced etch rate. To address this challenge, the mask windows are crafted with an asymmetrical layout - one side features a circular shape while the opposite side is designed as an elongated stripe. This asymmetry creates an imbalance in HF pressure, thereby directing the HF flow across the center, ensuring that the etch rate remains consistent throughout the structure. The vapor HF etch is carried out at a temperature 15 \degree C above room temperature (in our case, T = 315K) for 23 minutes, a condition intentionally set to achieve an optimal etch rate and minimal moisture concentration in the photoresist. After the vapor HF etch, the sample is carefully immersed in acetone and IPA to dissolve the photoresist layer. Finally, a critical point drying is performed. The IPA solution is replaced with liquid carbon dioxide, allowing the sample to dry while preserving the suspension of the lithium niobate structure.
\begin{table}[t]
\caption{Measrued and Extracted Parameters of Lithium Niobate SH0 Resonators.}
\label{table}
\setlength{\tabcolsep}{3pt}
\begin{tabular}{>{\centering}p{25pt}>{\centering}p{50pt}>{\centering}p{34pt}>{\centering}p{34pt}>{\centering}p{34pt}>{\centering\arraybackslash}p{34pt}}
\hline\hline
IDTs &  
Frequency & 
\(\mathrm{C_0}\) &
\(\mathrm{C_x}\) &
\(\mathrm{L_x}\) &
\(\mathrm{R_x}\)
\\
\hline
+9\degree{} & 
418 MHz & 
230 fF &
35.3 fF & 
4.11 \(\mathrm{\mu H}\) & 
5.75 \(\mathrm{\Omega}\) \\
\hline\hline
\end{tabular}
\label{tab:2}
\end{table}
\newline

\section{Experimental Results And Discussion}
\label{sec:results}

\subsection{SH0 mode and S0 mode}

\begin{figure}[t]
\centerline{\includegraphics[width=\columnwidth]{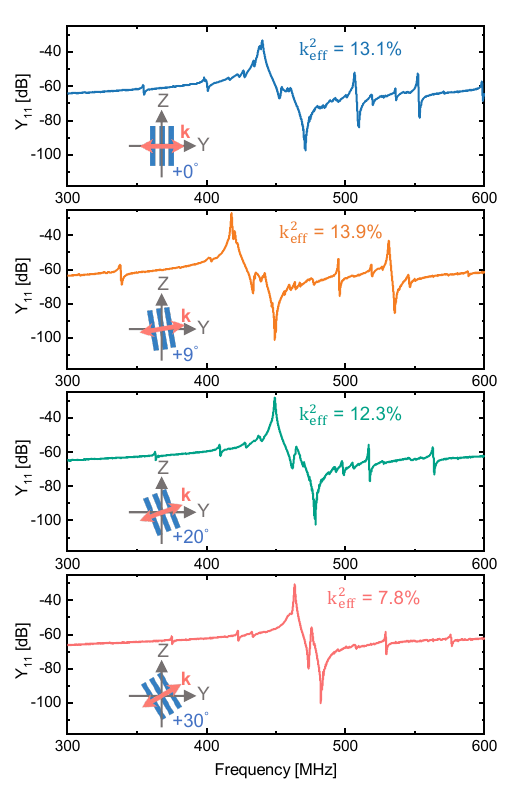}}
\caption{ RF response of resonators with IDTs aligned in different orientation.
}
\label{fig:7}
\end{figure}

The one-port scattering parameters of resonators are assessed using a network analyzer (Agilent PNA-L N5230A). Parasitic pad feed-through capacitance is canceled through de-embedding process\cite{pozar2011microwave}. The scanning range is set at 700 MHz, with 20,000 sampling points. A resolution bandwidth of 10 KHz is employed, and the input power is maintained at -12 dBm. All measurements are conducted in atmosphere. Measured S parameters are transformed into admittance (\(\mathrm{Y_{11}}\)). The Q is computed using the method decribed in \cite{4803696}. And effective coupling \(\mathrm{k_{eff}^2}\) is calculated as:
\(\mathrm{k_{eff}^2 = 2\times(f_p-f_s)/f_s}\), where \(\mathrm{f_p} \) represents parallel resonance and \(\mathrm{f_s} \) represents series resonance.

When IDTs are aligned from +0\degree{} to +40\degree{}, sheer horizontal modes (SH0) are detected. Different orders of SH0 modes emerge at varied frequencies between 350 MHz and 600 MHz. With 11-finger IDTs equipped, the \(\mathrm{10^{th}}\) order SH0 modes typically exhibit higher effective coupling \(\mathrm{k_{eff}^2}\) and Q. The edges are set at the zero points of displacement field, so the FP boundary conditions for best coupling SH0 modes are fulfilled. As shown in Fig. \ref{fig:6}, the maximum product of \(\mathrm{k_{eff}^2 \times Q}\) is observed when the IDTs are aligned at +9\degree{}, where \(\mathrm{10^{th}}\) order SH0 mode is found at 418 MHz. Q is 2117 and \(\mathrm{k_eff^2}\) reaches 13.9\%. Thus the figure of merit (FOM) \(\mathrm{k_{eff}^2 \times Q}\) is computed to be 294, one of the highest on LN-on-LN platform. The response of \(\mathrm{10^{th}}\) order SH0 mode is analyzed using Butterworth-Van Dyke (BVD) model illustrated in Fig. \ref{fig:6} (a). The extracted parameters \(\mathrm{C_0}\), \(\mathrm{C_x}\), \(\mathrm{L_x}\) and \(\mathrm{R_x}\) are listed in the Table \ref{tab:1}. The measurement results of \(\mathrm{10^{th}}\) order SH0 agree well with finite element simulation.  

\begin{figure}[t]
\centerline{\includegraphics[width=\columnwidth]{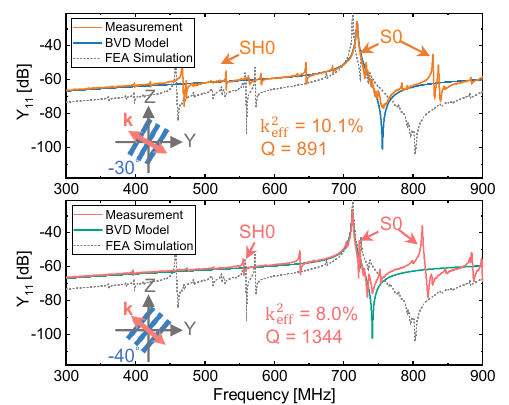}}
\caption{ 
 Measured results, BVD model analysis and finite  element analysis of S0 modes with -30\degree{} and -40\degree{} IDT orientation. 10 IDT fingers are equipped. Insignificant SH0 modes and higher order S0 modes are marked.
}
\label{fig:8}
\end{figure}

\begin{table}[t]
\caption{Measrued and Extracted Parameters of Lithium Niobate S0 Resonators.}
\label{table}
\setlength{\tabcolsep}{3pt}
\begin{tabular}{>{\centering}p{25pt}>{\centering}p{50pt}>{\centering}p{34pt}>{\centering}p{34pt}>{\centering}p{34pt}>{\centering\arraybackslash}p{34pt}}
\hline\hline
IDTs &  
Frequency & 
\(\mathrm{C_0}\) &
\(\mathrm{C_x}\) &
\(\mathrm{L_x}\) &
\(\mathrm{R_x}\)
\\
\hline
-30\degree{} & 
719 MHz & 
216 fF &
22.5 fF & 
2.18 \(\mathrm{\mu H}\) & 
8.55 \(\mathrm{\Omega}\) \\
\hline
-40\degree & 
713 MHz & 
216 fF &
17.7 fF & 
2.81 \(\mathrm{\mu H}\) & 
7.70 \(\mathrm{\Omega}\)
\\
\hline\hline
\end{tabular}
\label{tab:3}
\end{table}

The substantial influence of the frequency response of SH0 modes on IDT orientation is confirmed experimentally. As the IDT fingers move away from +9\degree{} to +0\degree{}, +20\degree{}, and +30\degree{}, we observe a decrease in coupling and an increase in resonant frequency, as shown in Fig. \ref{fig:7}. These findings align with our simulation results and can be attributed to the anisotropy present in the LN piezoelectric thin film. As the IDTs' alignment deviates significantly from +9\degree{}, the SH0 strength becomes less pronounced. However, maximum coupling of S0 mode is achieved when IDTs are aligned from -30\degree{} to -40\degree{}, as demonstrated in Fig. \ref{fig:8}. At a -30\degree{} IDT orientation, the most prominent S0 mode emerges at 719 MHz with an effective coupling of 10.1\% (\(\mathrm{k_{eff}^2}\)) and Q = 891. At a -40\degree{} IDT orientation, the most noticeable S0 mode appears at 713 MHz with 8.0\% \(\mathrm{k_{eff}^2}\) coupling and Q = 1344. We analyzed the response of S0 modes using the BVD model and compared the results with COMSOL finite element simulations. The extracted parameters are provided in Table \ref{tab:2}. It is worth noting that S0 modes are highly sensitive to the resonators' boundary conditions\cite{7005450}. Maximum coupling and Q can only be obtained when the edge of resonator is positioned at the stress field nodes, which is different from the boundary conditioned required by SH0 modes. Fabrication errors could affect the performance of the S0 mode resonator. In our case, although the resonant frequency aligns with the simulation results, the coupling coefficient decreases because the aforementioned FP boundary conditions are not fulfilled for S0 modes. Even slight mismatch between nodes of S0 mode and IDT finger configuration could impact the coupling.
\begin{figure}[t]
\centerline{\includegraphics[width=\columnwidth]{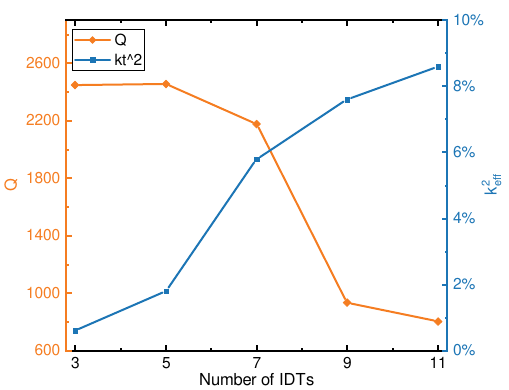}}
\caption{ 
 Measured Q and \(\mathrm{k_{eff}^2}\) of S0 modes with -30\degree{} IDTs and different number of IDT fingers.
}
\label{fig:9}
\end{figure}

\subsection{Electrode Loading}

Despite aluminum's low inherent loss and superior acoustic impedance compatibility with LN, it offers lower conductivity compared to unreactive metals such as gold. To delve deeper into the impact of electrode loading on Q, we fabricated identical S0/SH0 mode resonator structures, maintaining the same dimensions but varying the number of IDT fingers. Our measurements suggest that S0 modes are more vulnerable to electrode loading. As shown in Fig. \ref{fig:9}, when the IDT fingers are aligned at -30\degree{} relative to the z-axis, reducing the number of IDT fingers from 11 to 3 leads to a decrease in energy coupled to the acoustic regime from RF regime, which is represented by the effective coupling \(\mathrm{k_{eff}^2}\) dropping from 8.6\% to 0.6\%, while Q elevates from 801 to over 2400. This implies a significant energy loss in S0 modes due to electrode loading. In contrast, for the SH0 mode resonator, electrode loading has a lesser impact. With a +9\degree{} IDT orientation and an increase in the number of IDTs from 3 to 11, we observe an increase in \(\mathrm{k_{eff}^2}\) while Q remains consistently high, exceeding 1500.

\begin{figure}[t]
\centerline{\includegraphics[width=\columnwidth]{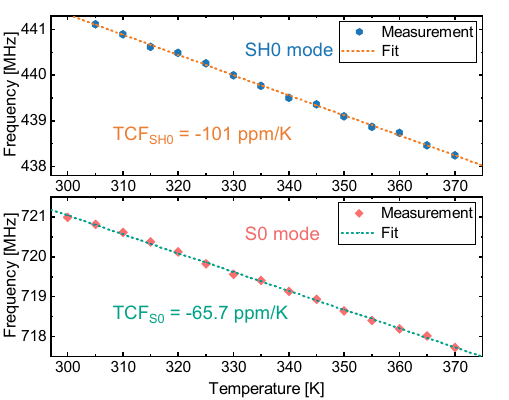}}
\caption{ 
 Measured Temperature response of S0 modes and SH0 modes
}
\label{fig:10}
\end{figure}

\subsection{Temperature Stability}

We examined the temperature coefficient (TCF) of LN resonators of various orientations in their fundamental mode by incrementally adjusting the temperature from 300K to 370K in steps of 5K. Each measurement was conducted after allowing a 5-minute stabilization period at the desired temperature. The results, depicted in Fig. \ref{fig:10}, reveal that both SH0 and S0 modes display a highly linear and stable TCF. The TCF for SH0 modes, recorded at -101 ppm/K, is slightly higher than that for S0 modes, which stands at -65.7 ppm/K. These TCF values align with those observed in uncompensated LN SAW devices\cite{4803491}. Compensation for temperature can be achieved by incorporating an additional \(\mathrm{SiO_2}\) layer, as explored in previous studies\cite{6782639, 7863565, 7329342}.

\subsection{Nonlinearity and Power Handling}

\begin{figure}[t]
\centerline{\includegraphics[width=\columnwidth]{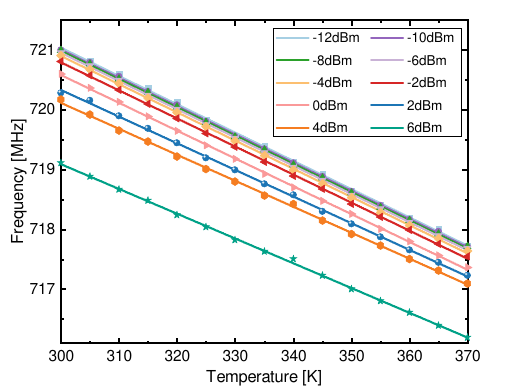}}
\caption{ 
 Measured S0 mode Temperature response at different power level.
}
\label{fig:11}
\end{figure}

\begin{figure}[t]
\centerline{\includegraphics[width=\columnwidth]{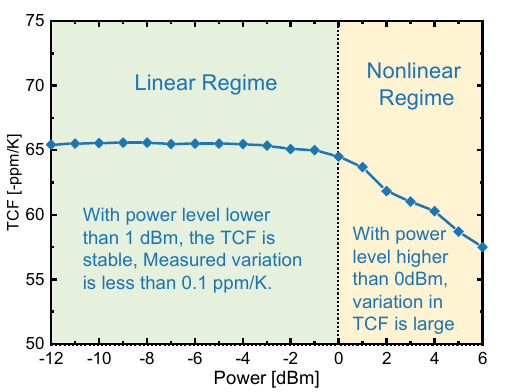}}
\caption{ 
 Measured TCF at different power level.
}
\label{fig:12}
\end{figure}

We conducted an empirical investigation into the nonlinearity behavior of LN resonators by measuring the frequency response across various power levels. An S0 mode resonator with 11 IDT orientations -30\degree{} from the +z axis, and dimensions of 70 \(\mathrm{\mu m \times}\) 44 \( \mathrm{\mu m} \), was subjected to a frequency sweep from 680MHz to 790MHz at input power levels ranging from -12 dBm to 10 dBm, incremented by 1 dBm. The device operated within its linear region, as evidenced by a Lorentzian-shaped frequency response near resonance when the input power was below -1 dBm. As the input power increased beyond this level, a progressively larger resonant frequency shift and distortion of frequency response became evident, signifying the transition of the resonator into its non-linear regime. A bifurcation point was identified at 6 dBm, marking the threshold of instability.

We investigated the power handling capacity of our LN-on-LN resonator devices by conducting a frequency sweep at varying temperature ranges from 300K to 370K and power levels spanning from -12 dBm to 6 dBm. Resonant frequencies of our LN-on-LN devices under these varied conditions are displayed in Fig. \ref{fig:11}. It is discernible that for power levels below 0 dBm, the shift in resonant frequencies remains minimal. However, upon exceeding a power level of 0 dBm, substantial frequency shifts are observable, even though the temperature response maintains its linearity. Further study of the TCF at various power levels, depicted in Fig. \ref{fig:12}, shows that the TCF variation remains under 0.1 ppm/K for power levels less than 0 dBm, suggesting a stable temperature response. Contrastingly, SAW LN-on-Si devices exhibited significant TCF alterations even at lower power levels \cite{9385101}. We attribute the stability of our measured TCF in relation to input power to the minimization of residue stress in the LN-on-LN thin film resulting from thermal cycling induced by power in the matched substrate. 

\section{Conclusion}

In this work, we pioneered the design, fabrication, and characterization of first-ever laterally vibrating SH0 and S0 mode resonators on an LN-on-LN platform driven by aluminum-driven IDTs. This is made possible by the development of a novel fabrication methodology compatible with aluminum. Incorporating aluminum electrodes, recognized for their low mechanical loss, enabled us to demonstrate a quality factor (Q) enhancement exceeding 2000 and high effective electromechanical coupling coefficient (\(\mathrm{k_{eff}^2}\)) surpassing 13\%. Our work yielded a figure of merit (FOM) of 294, one of the highest within the same platform and competitive with LN-on-Si outcomes. However, our LN-on-LN devices still maintain several advantages over traditional LN-on-Si devices. In contrast with the limits on the thermal instablity of LN-on-Si platform, we demonstrated a linear and consistent temperature coefficient of frequency (TCF) across different temperatures. Furthermore, our devices exhibit remarkable sensitivity to input power in terms of oscillation frequency, with less than 0.1 ppm/K within the resonator's linear range. Such traits potentially broaden the applications of LN-on-LN devices, including the utilization in highly sensitive, uncooled sensors facilitated by integrated resonator arrays on a monolithic chip\cite{7050889, niklaus2008mems, gokhale2013uncooled}.

\section*{Acknowledgment}

The devices were fabricated and tested in Birck Nanotechnology Center at Purdue University. The authors would like to thank Hao Tian, Noah Opondo and Ozan Erturk for valuable discussion on fabrication techniques, Mengyue Xu for discussion on lithium niobate devices simulation, Neil Dilley for assistance in handling of AJA ion milling tool and Nithin Raghunathan for training on vapor HF tool. 

\bibliographystyle{IEEEtran}

\bibliography{}

\begin{IEEEbiography}[{\includegraphics[width=1in,height=1.25in,clip,keepaspectratio]{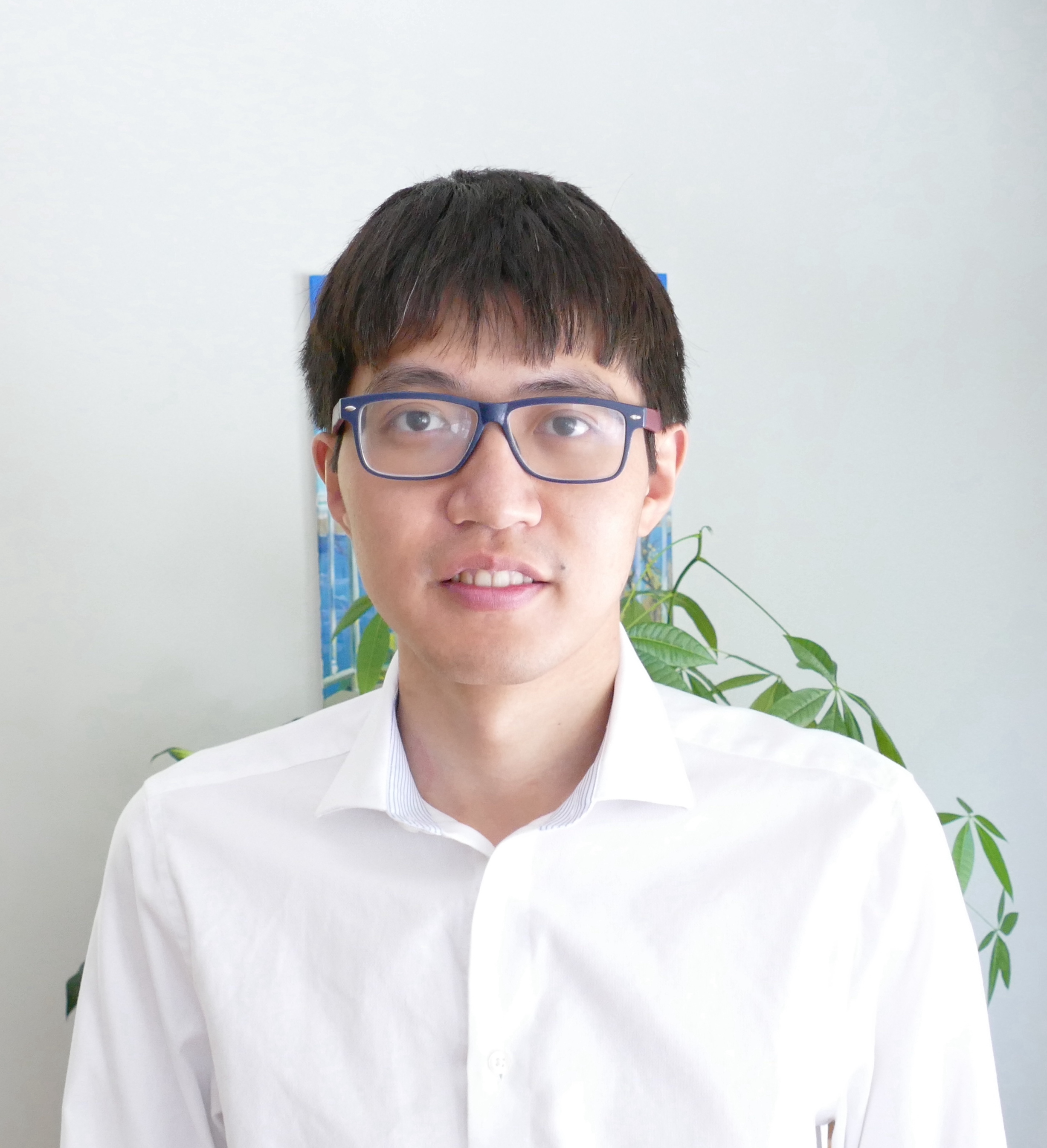}}]{Yiyang Feng} 
received the B.Sc. degree in physics from Peking University, Beijing, China, in 2016. He is currently pursuing the Ph.D. degree with the Department of Physics and Astronomy, Purdue University, West Lafayette, IN, USA. His research focuses on magnetostatic wave devices and niobate acoustic devices and their RF applications.
\end{IEEEbiography}

\begin{IEEEbiography}[{\includegraphics[width=1in,height=1.25in,clip,keepaspectratio]{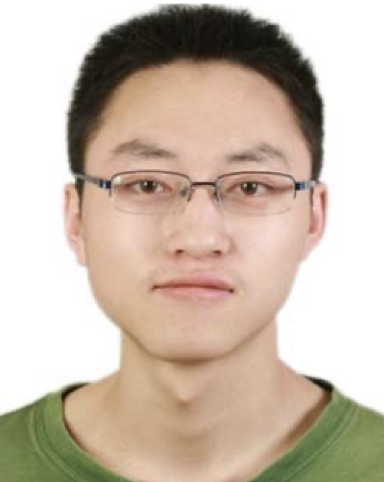}}]{Sen Dai} (member, IEEE) received the B.Sc. degree in physics from the University of Science and Technology of China, Hefei, Anhui, China, in 2014. He joined Prof. Sunil Bhave’s OxideMEMS Lab in January 2017. His research is focused on RF MEMS resonators and micromachining ferrite components. He received Ph.D. degree from the Department of Physics and Astronomy, Purdue University, West Lafayette, IN, USA in 2021. After graduation, he joined Qorvo, Apopka, FL, USA.
\end{IEEEbiography}

\begin{IEEEbiography}[{\includegraphics[width=1in,height=1.25in,clip,keepaspectratio]{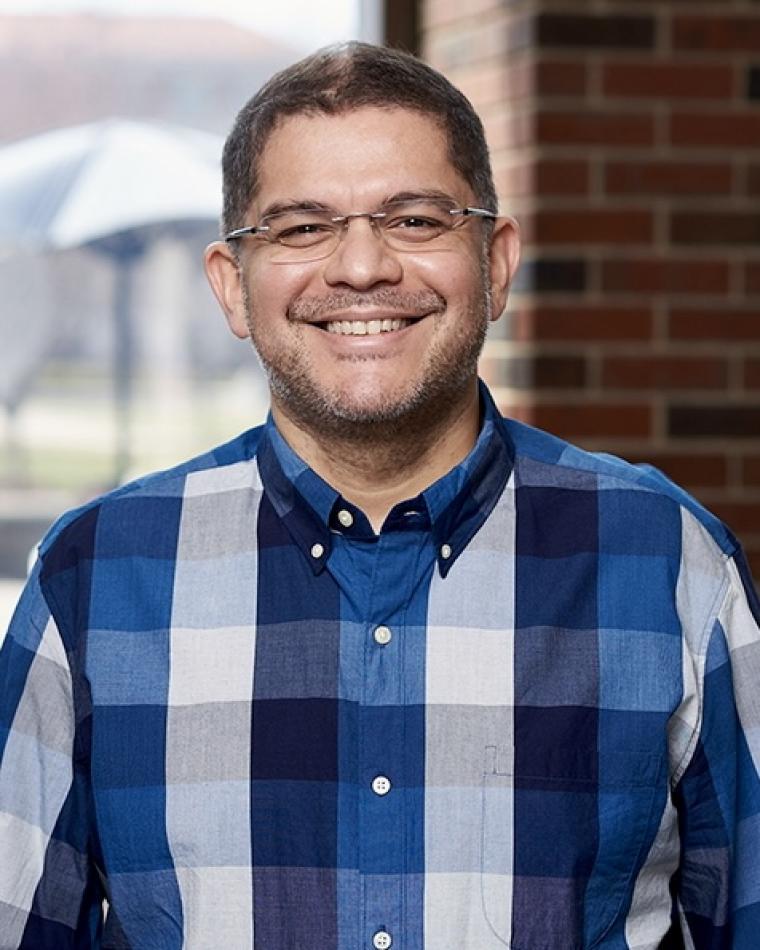}}]{Sunil A. Bhave} (Senior Member, IEEE) received the B.S. and Ph.D. degrees in electrical engineering and computer sciences from the University of California at Berkeley, Berkeley, CA, USA, in 1998 and 2004, respectively.
He was a Professor with Cornell University, Ithaca, NY, USA, for ten years and worked at Analog Devices, Woburn, MA, USA, for five years. In April 2015, he joined the Department of Electrical and Computer Engineering, Purdue University, West Lafayette, IN, USA, where
he is currently the Associate Director of operations at the Birck Nanotechnology Center. He is a Co-Founder of Silicon Clocks, Fremont, CA, USA, that was acquired by Silicon Labs, Austin, TX, USA, in April 2010. His research interests focus on the interdomain coupling in optomechanical, spin-acoustic, and color center-MEMS devices.
Dr. Bhave received the NSF CAREER Award in 2007, the DARPA Young Faculty Award in 2008, the IEEE Ultrasonics Society’s Young Investigator Award in 2014, and the Google Faculty Research Award in 2020. His students have received best paper awards at the IEEE Photonics 2012, the IEEE Ultrasonics Symposium 2009, and IEDM 2007.
\end{IEEEbiography}

\end{document}